\documentclass[preprint,aps]{revtex4}
\newcommand{\be}{\begin{equation}}
\newcommand{\ee}{\end{equation}}
\newcommand{\bea}{\begin{eqnarray}}
\newcommand{\eea}{\end{eqnarray}}
\newcommand{\ba}{\begin{array}}
\newcommand{\ea}{\end{array}}

\begin{document}
\title{Strangeness Content of the Nucleon in the $\chi$CQM$_{{\rm config}}$}
\author{Harleen Dahiya  and Manmohan Gupta}
\affiliation{ Department of Physics, Centre of Advanced Study in
Physics,\\ Panjab University, Chandigarh 160014, India}
\date{\today}

\begin{abstract}
Several parameters characterizing the strangeness content of the
nucleon have been calculated in the chiral constituent quark model
with configuration mixing ($\chi$CQM$_{{\rm config}}$) which is
known to provide a satisfactory explanation of the ``proton spin
problem'' and related issues. In particular, we have calculated
the strange spin polarization $\Delta s$, the strangeness
contribution to the weak axial vector couplings $\Delta_8$ etc.,
strangeness contribution to the magnetic moments $\mu(p)^s$ etc.,
the strange quark flavor fraction $f_s$, the strangeness dependent
quark ratios $\frac{2 \bar s}{u+d}$ and $\frac{2 \bar s}{\bar u+
\bar d}$ etc.. Our results show in general excellent agreement
with the recent experimental observations.

\end{abstract}

\maketitle

The recent measurements by  MIT-Bates (SAMPLE) \cite{sample} and
by JLab (HAPPEX) \cite{happex} regarding the contribution of
strangeness to the magnetic moment of the proton ($\mu(p)^s$) have
triggered a great deal of interest \cite{mup,mup-aw} in finding
the strangeness content of the nucleon. These experiment have
respectively observed $\mu(p)^s$ to be $-0.36 \pm 0.20$
\cite{sample} and $-0.038 \pm 0.042$ \cite{happex}, predicted to
be zero in naive constituent quark model (CQM)
\cite{dgg,Isgur,yaouanc}. The broader question of strangeness
content of the nucleon has also been discussed by several authors
recently \cite{ellis-strange}. It is widely recognized that a
knowledge about the strangeness content of the nucleon would
undoubtedly provide vital clues to the non-perturbative aspects of
QCD.

The existence of strangeness in the nucleon has been indicated in
the context of low energy experiments \cite{ao,nutev,sigma}
whereas it has been observed in the deep inelastic scattering
(DIS) experiments \cite{emc,smc,adams}. In the context of DIS, the
strange quark polarized structure functions of the nucleon
\cite{ellis-kar} looks to be well established through the
measurements of polarized structure functions of the nucleon. The
present experimental situation \cite{smc,adams} in terms of the
strange spin polarization is summarized as $\Delta s=-0.10 \pm
0.04$ \cite{smc}, $-0.07 \pm 0.03$ \cite{adams}. Apart from the
observations of DIS data regarding strangeness dependent spin
polarization functions, several interesting facts have also been
revealed regarding the quark flavor distribution functions in the
DIS experiments \cite{e866} indicating that the flavor structure
of the nucleon is not limited to $u$ and $d$ quarks only. In
particular, the CCFR Collaboration in their neutrino charm
production experiment \cite{ao} has given fairly good deal of
information regarding the strangeness dependent quark ratios in
the nucleon. This has recently been confirmed by the NuTeV
collaboration \cite{nutev} with greater accuracy and the ratios
are given as $\frac{2 \bar s}{u+d} =0.099^{+0.009}_{-0.006}$ and
$\frac{2\bar s}{\bar u+\bar d} =0.477^{+0.063}_{-0.053}.$ In the
context of low energy experiments, the large pion-nucleon sigma
term value \cite{sigma} indicating non zero strange quark flavor
fraction $f_s$ is also indicative of the presence of strange
quarks in the nucleon although there is no consensus regarding the
various mechanisms which can contribute to  $f_s$
\cite{cheng,gasser,scadron}. Therefore, the indications of the
strange quark degree of freedom in DIS as well as low energy
experiments provide a strong motivation to examine the strangeness
contribution to the nucleon thereby giving vital clues to the
non-perturbative effects of QCD.

One may think that the strangeness content of the nucleon perhaphs
can be obtained through the generation of ``quark sea''
perturbatively from the quark-pair production by gluons. However,
this kind of ``sea'' is symmetric w.r.t. $\bar u$ and $\bar d$
\cite{cheng}, negated by the observed value of $\bar u-\bar d$
asymmetry \cite{e866}. Therefore, one has to consider the ``quark
sea'' produced by the non-perturbative mechanism. One such model
which can yield an adequate description of the ``quark sea''
generation through the chiral fluctuations is the chiral
constituent quark model ($\chi$CQM) \cite{cheng,manohar} which is
not only successful in giving a satisfactory explanation of
``proton spin crisis'' \cite{emc} but is also able to account for
the violation of Gottfried Sum Rule {\cite{{e866},{GSR}}}, baryon
magnetic moments and hyperon $\beta-$decay parameters
\cite{cheng}. Recently, it has been shown that configuration
mixing generated by spin-spin forces improves the predictions of
$\chi$CQM regarding the quark distribution functions and spin
polarization functions \cite{hd}. Further, the chiral constituent
quark model with configuration mixing ($\chi$CQM$_{{\rm config}}$)
when coupled with the quark sea polarization and orbital angular
momentum through the Cheng-Li mechansim \cite{cheng} is able to
give an excellent fit \cite{hdorbit} to the octet magnetic
moments. It, therefore, becomes desirable to carry out a detailed
analysis in the $\chi$CQM$_{{\rm config}}$ of the strangeness
dependent spin polarization functions as well as the quark
distribution functions, particularly in the light of some recent
observations \cite{sample,happex,ao,nutev,sigma,adams,bass}.

The purpose of the present communication is to carry out detailed
calculations of the parameters  characterizing the strangeness of
the nucleon within the $\chi$CQM$_{{\rm config}}$. In particular,
we would like to calculate the strange spin polarization $\Delta
s$, strangeness contribution to the weak axial vector couplings
$\Delta_3$, $\Delta_8$ and $\Delta_0$, strangeness contribution to
the magnetic moments $\mu(p)^s$ and $\mu(n)^s$,  the strange quark
flavor fraction $f_s$,  the strangeness dependent quark ratios
$\frac{2 \bar s}{u+d}$ and $\frac{2 \bar s}{\bar u+ \bar d}$. For
the sake of completeness, we would also like to calculate the
strangeness contribution to the magnetic moments of decuplet
baryons $\mu(\Delta^{++})^s$, $\mu(\Delta^{+})^s$,
$\mu(\Delta^{o})^s$ and $\mu(\Delta^{-})^s$ which have not been
observed experimentally.

To make the mss. more readable as well as for ready reference, we
mention the essentials of $\chi$CQM$_{{\rm config}}$, for details
we refer the reader to \cite{cheng,hd}. The basic process in the
$\chi$CQM formalism is the emission of a Goldstone boson (GB) by a
constituent quark which further splits into a $q \bar q$ pair, for
example,

\be
  q_{\pm} \rightarrow {\rm GB}^{0}
  + q^{'}_{\mp} \rightarrow  (q \bar q^{'})
  +q_{\mp}^{'}\,,                              \label{basic}
\ee where $q \bar q^{'} +q^{'}$ constitute the ``quark sea''
\cite{cheng} and the $\pm$ signs refer to the quark helicities.
The effective Lagrangian describing interaction between quarks and
a nonet of GBs, consisting of octet and a singlet, can be
expressed as
\be
{\cal L}= g_8 {\bf \bar q}\left(\Phi+\zeta\frac{\eta'}{\sqrt 3}I
\right) {\bf q}=g_8 {\bf \bar q}\left(\Phi'\right) {\bf q}\,, \ee
where  $\zeta=g_1/g_8$, $g_1$ and $g_8$ are the coupling constants
for the singlet and octet GBs, respectively, $I$ is the $3\times
3$ identity matrix. The GB field which includes the octet and the
singlet GBs is written as \bea
 \Phi' = \left( \ba{ccc} \frac{\pi^0}{\sqrt 2}
+\beta\frac{\eta}{\sqrt 6}+\zeta\frac{\eta^{'}}{\sqrt 3} & \pi^+
  & \alpha K^+   \\
\pi^- & -\frac{\pi^0}{\sqrt 2} +\beta \frac{\eta}{\sqrt 6}
+\zeta\frac{\eta^{'}}{\sqrt 3}  &  \alpha K^0  \\
 \alpha K^-  &  \alpha \bar{K}^0  &  -\beta \frac{2\eta}{\sqrt 6}
 +\zeta\frac{\eta^{'}}{\sqrt 3} \ea \right) {\rm and} ~~~~q =\left( \ba{c} u \\ d \\ s \ea
\right)\,. \eea

SU(3) symmetry breaking is introduced by considering $M_s >
M_{u,d}$ as well as by considering the masses of GBs to be
nondegenerate
 $(M_{K,\eta} > M_{\pi}$ and $M_{\eta^{'}} > M_{K,\eta})$
\cite{cheng}. The parameter $a(=|g_8|^2$) denotes the probability
of chiral fluctuation  $u(d) \rightarrow d(u) + \pi^{+(-)}$,
$\alpha^2 a$, $\beta^2 a$ and $\zeta^2 a$
 respectively denote the probabilities of fluctuations
$u(d) \rightarrow s + K^{-(0)}$,  $u(d,s) \rightarrow u(d,s) +
\eta$ and $u(d,s) \rightarrow u(d,s) + \eta^{'}$.  Further, to
make the transition from $\chi$CQM to $\chi$CQM$_{{\rm config}}$,
the nucleon wavefunction gets modified because of the
configuration mixing generated by the chromodynamic spin-spin
forces \cite{{dgg},{Isgur},{yaouanc},hd} as follows,
\begin{equation}
|B\rangle = \cos \phi |56,0^+\rangle_{N=0} + \sin
\phi|70,0^+\rangle_{N=2}\,, \label{mixed}
\end{equation}
where $\phi$ represents the $|56\rangle-|70\rangle$ mixing. For
details of the  spin, isospin and spatial parts of the
wavefunction,  we  refer the reader to {\cite{{yaoubook}}.

Before propceeding further, we briefly discuss the strangeness
dependent spin polarization functions, quark distribution
functions and the related quantities of the nucleon. To begin
with, we consider the spin structure of a nucleon defined as
\cite{{cheng}}
\be
\hat B \equiv \langle B|N|B\rangle, \ee where $|B\rangle$ is the
nucleon wavefunction defined in Eq. (\ref{mixed}) and $N$ is the
number operator given by
\be
 N=n_{u_{+}}u_{+} + n_{u_{-}}u_{-} +
n_{d_{+}}d_{+} + n_{d_{-}}d_{-} + n_{s_{+}}s_{+} +
n_{s_{-}}s_{-}\,, \ee  $n_{q_{\pm}}$ being the number of $q_{\pm}$
quarks. Following Ref. \cite{hd}, the contribution to the proton
spin by different quark flavors in $\chi$CQM$_{{\rm config}}$ can
be given by the spin polarizations $\Delta q=q_+-q_-$ expressed as
\be
   \Delta u =\cos^2 \phi \left[\frac{4}{3}-\frac{a}{3}
   (7+4 \alpha^2+ \frac{4}{3} \beta^2
   + \frac{8}{3} \zeta^2)\right]+ \sin^2 \phi \left[\frac{2}{3}-\frac{a}{3} (5+2 \alpha^2+
  \frac{2}{3} \beta^2 + \frac{4}{3} \zeta^2)\right], \label {du}\ee
\be
  \Delta d =\cos^2 \phi \left[-\frac{1}{3}-\frac{a}{3} (2-\alpha^2-
  \frac{1}{3}\beta^2- \frac{2}{3} \zeta^2)\right]
+ \sin^2 \phi \left[\frac{1}{3}-\frac{a}{3} (4+\alpha^2+
  \frac{1}{3} \beta^2 + \frac{2}{3} \zeta^2)\right], \label{dd}
  \ee
\be   \Delta s =\cos^2 \phi \left[ -a \alpha^2 \right]+\sin^2 \phi
\left[ -a \alpha^2 \right]\,. \ee A closer look at the above
equations reveals that $\phi$ represents the configuration mixing
angle, the constant factors represent the CQM results and the
factors which are multiple of $a$ represent the contribution from
the ``quark sea''. It is important to mention here that the
presence of $s \bar s$ in the ``quark sea'' (Eq. (\ref{basic}))
involves the terms $\alpha^2 a$, $\beta^2 a$ and $\zeta^2 a$
respectively denoting the fluctuations $u(d) \rightarrow s +
K^{+(0)}$, $u(d,s) \rightarrow u(d,s) + \eta$ and $u(d,s)
\rightarrow u(d,s) + \eta^{'}$.  It is clear from the above
fluctuations that the strange quarks come from the fluctuations of
the $u$ and $d$ quarks as well and therefore, apart from
contributing to the strange spin polarization, the strange quarks
also contribute to the spin polarizations of $u$ and $d$ quarks.
It should also be noted that the $\bar d-\bar u$ asymmetry in this
case can be easily understood in terms of the above fluctuations
where the valence $u$ quarks are likely to produce more $\bar d$
than the valence $d$ quarks producing $\bar u$ in the proton. The
spin polarization functions are related to the axial vector
couplings measured in the baryon weak decays \cite{ellis-kar}, for
example, the non-singlet combinations of the quark spin
polarizations ($\Delta_3$ and $\Delta_8$) can be expressed as \bea
\Delta_3&=& \Delta u-\Delta d\,=F+D\,,
\\ \Delta_8&=& \Delta u+\Delta d-2 \Delta s=3F-D\,, \eea
where $F$ and $D$ are the usual SU(3) parameters characterizing
the weak matrix elements. The flavor singlet combination on the
other hand can be related to the total spin carried by the quarks
as \bea \Delta_0&=&\frac{1}{2} \Delta \Sigma= \frac{1}{2}(\Delta
u+\Delta d+\Delta s)\,. \eea

After discussing the spin polarization functions, we would like to
briefly discuss the formalism for the strangeness contribution to
the magnetic moments. The magnetic moment of a given baryon in the
$\chi$CQM can be expressed as
\be
\mu(B)_{{\rm total}} = \mu(B)_{{\rm val}} + \mu(B)_{{\rm sea}}\,,
\ee where $\mu(B)_{{\rm val}}$ represents the contribution of the
valence quarks and $\mu(B)_{{\rm sea}}$ corresponding to the
``quark sea'' (Eq. (\ref{basic})). Further, $\mu(B)_{{\rm sea}}$
can be written as
\be
\mu(B)_{{\rm sea}} = \mu(B)_{{\rm spin}} + \mu(B)_{{\rm orbit}}\,,
\ee where the first term is the magnetic moment contribution of
the $q^{'}$ in Eq. (\ref{basic}) and the second term is due to the
rotational motion of the two bodies, $q^{'}$ and GB, corresponding
to the fluctuation given in Eq. (\ref{basic}) as proposed by Cheng
and Li  \cite{cheng}.

To find the strangeness contribution to the magnetic moment of the
proton $\mu(p)^s$ we should note that there are no `strange'
valence quarks, therefore $\mu(p)^s$ receives contributions only
from the ``quark sea" and is expressed as
\be
\mu(p)^s = \mu(p)^s_{{\rm spin}} + \mu(p)^s_{{\rm orbit}}\,.
\label{totalmag} \ee Following Ref. \cite{hdorbit},
$\mu(p)^s_{{\rm spin}}$ in the chiral constituent quark model with
configuration mixing can be expressed as
\be
\mu(p)^s_{{\rm spin}}=\sum_{q=u,d,s}\Delta q(p)^s_{{\rm sea}}\mu_q
\,, \label{magsea} \ee
where $\mu_q (q=u,d,s)$ being the quark magnetic moment and
\be
   \Delta u(p)^s_{{\rm sea}} =\cos^2 \phi \left[-\frac{a}{3}
   (4 \alpha^2+ \frac{4}{3} \beta^2
   + \frac{8}{3} \zeta^2)\right]+ \sin^2 \phi \left[-\frac{a}{3} (2 \alpha^2+
  \frac{2}{3} \beta^2 + \frac{4}{3} \zeta^2)\right], \label{dusea}\ee
\be
  \Delta d(p)^s_{{\rm sea}}=\cos^2 \phi \left[-\frac{a}{3} (-\alpha^2-
  \frac{1}{3}\beta^2- \frac{2}{3} \zeta^2)\right]
+ \sin^2 \phi \left[-\frac{a}{3} (\alpha^2+
  \frac{1}{3} \beta^2 + \frac{2}{3} \zeta^2)\right],
  \ee
\be  \Delta s(p)^s_{{\rm sea}} =\cos^2 \phi \left[ -a \alpha^2
\right]+\sin^2 \phi \left[ -a \alpha^2 \right]\,. \label{dssea}
\ee
Similarly, the ``quark sea'' contribution to the orbital angular
momentum is expressed as
\be
\mu(p)^s_{{\rm orbit}} = \frac{4}{3} [\mu (u_+ \rightarrow s_-)]-
\frac{1}{3} [\mu (d_+ \rightarrow s_-)]\,, \label{orbit p} \ee
where
\be
\mu(q_+ \rightarrow s_-) =\frac{e_{s}}{2M_q}\langle
l_q\rangle+\frac{e_{q}-e_{s}}{2M_{GB}}\langle l_{GB}\rangle\,. \ee
The quantities $(l_q, l_{GB})$ and $(M_q, M_{GB})$ are the orbital
angular momenta and masses of quark and GB, respectively. The
orbital angular momentum contribution to the magnetic moment due
to all the fluctuations is then given as \bea \mu(u_+ \rightarrow
s_-) &=& a \left[ -\frac{M^2_{\pi}}{2
{M}_{\pi}(M_u+{M}_{\pi})}-\frac{\alpha^2(M^2_{K}-3M^2_u)}{2
{M}_{K}(M_u+{M}_{K})}+\frac{(3+\beta^2+2 \zeta^2)M^2_{\eta}}{6
{M}_{\eta}(M_u+{M}_{\eta})} \right]{\mu}_N, \label{u} \\ \mu(d_+
\rightarrow s_-) &=& a \frac{M_u}{M_d} \left[ -\frac{\alpha^2
M^2_{K}}{2 {M}_{K}(M_d+{M}_{K})}-\frac{(\beta^2+2
\zeta^2)M^2_{\eta}}{12 {M}_{\eta}(M_d+{M}_{\eta})} \right]{\mu}_N,
\label{d}  \eea where  $M_{\pi}$, $M_{K}$ and $M_{\eta}$ are the
masses of pion, Kaon and $\eta$ respectively and $\mu_N$ is the
Bohr magneton. The strangeness contribution to the magnetic
moments of the neutron $n(ddu)$ as well as the decuplet baryons
$\Delta^{++}(uuu)$, $\Delta^{+}(uud)$, $\Delta^{o}(udd)$ and
$\Delta^{-}(ddd)$ can be calculated similarly.

The quark distribution functions incorporating the strangeness
content in the $\chi$CQM are expressed as \cite{{cheng},hd} \be
\bar u =\frac{1}{12}[(2 \zeta+\beta+1)^2 +20] a\,,~~~~  \bar d
=\frac{1}{12}[(2 \zeta+ \beta -1)^2 +32] a\,,~~~~
 \bar s =\frac{1}{3}[(\zeta -\beta)^2 +9 {\alpha}^{2}] a\,, \label{barq}\ee
 \be
  u-\bar u=2\,,~~~~~d-\bar d=1\,,~~~~~s-\bar s=0\,.
 \ee

Similarly, the other important quantities having implications for
the strangeness contribution to the nucleon are the quark flavor
fractions $f_q=\frac{q+\bar q}{\sum_{q} (q+\bar q)}$ which are
expressed in terms of the $\chi$CQM parameters as \bea f_u&=&
\frac{12+a(21+\beta^2+4 \zeta+4 \zeta^2+\beta(2+4
\zeta))}{3(6+a(9+\beta^2+6 \alpha^2+2 \zeta^2))}\,, \nonumber \\
f_d&=& \frac{6+a(33+\beta^2-4 \zeta+4 \zeta^2+\beta(-2+4
\zeta))}{3(6+a(9+\beta^2+6 \alpha^2+2 \zeta^2))}\,, \nonumber \\
 f_s&=& \frac{4a(\beta^2++9 \alpha^2-2
\beta\zeta+\zeta^2)}{3(6+a(9+\beta^2+6 \alpha^2+2 \zeta^2))}\,.
\label{fuds} \eea It is clear from the above expressions that the
non zero value of the parameters $a$, $\alpha$, $\beta$ and
$\zeta$ implies $f_s\neq 0$ as well as modify $f_u$ and $f_d$ due
to the strangeness contributions coming from the ``quark sea''.
Further, the ratio of the functions
\be f_3= f_u-f_d\,, ~~~~f_8= f_u+f_d-2 f_s \,, \ee  and the ratios
\be \frac{2 \bar s}{(u+d)}= \frac{4 a(9 \alpha^2 +\beta^2-2 \beta
\zeta +\zeta^2)}{18+a(27+\beta^2+4 \beta\zeta+4 \zeta^2)}\,, ~~~
\frac{2 \bar s}{(\bar u+\bar d)}= \frac{4(9 \alpha^2 +\beta^2-2
\beta \zeta +\zeta^2)}{27+\beta^2+4 \beta\zeta+4 \zeta^2}\,, \ee
have also been measured, therefore providing an opportunity to
check the strange quark content of the nucleon.

The $\chi$CQM$_{{\rm config}}$ involves five parameters, four of
these $a$, $a \alpha^2$, $a \beta^2$, $a \zeta^2$ representing
respectively the probabilities of fluctuations to pions, $K$,
$\eta$, $\eta^{'}$, following the hierarchy $a > \alpha > \beta >
\zeta$, while the fifth representing the mixing angle. The mixing
angle $\phi$ is fixed from the consideration of neutron charge
radius \cite{yaouanc}, whereas for the other parameters we would
like to update our analysis using the latest data \cite{PDG}. In
this context, we find it convenient to use $\Delta u$, $\Delta_3$,
$\bar u-\bar d$ and $\bar u/\bar d$ as inputs with their latest
values given in Tables \ref{spin} and \ref{quark}. Before carrying
out the fit to the above mentioned parameters, we would like to
find their ranges by qualitative arguments. To this end, the range
of the symmetry breaking parameter $a$ can be easily found by
considering the spin polarization function $\Delta u$, by giving
the full variation to the parameters $\alpha$, $\beta$ and
$\zeta$, for example, one finds $0.10 \lesssim a \lesssim 0.14$.
The range of the parameter $\zeta$ can be found from  $\bar u/\bar
d$ using the latest experimental measurement \cite{e866} and it
comes out to be $-0.70 \lesssim \zeta \lesssim -0.10$. Using the
above found ranges of $a$ and $\zeta$ as well as the latest
measurement of $\bar u-\bar d$ asymmetry \cite{e866}, $\beta$
comes out to be in the range $0.2\lesssim \beta \lesssim 0.7$.
Similarly, the range of $\alpha$ can be found by considering the
flavor non-singlet component $\Delta_3$ and it comes out to be
$0.2 \lesssim \alpha \lesssim 0.5$. After finding the ranges of
the symmetry breaking parameters, we have carried out a fine
grained analysis using the above ranges as well as considering
$\alpha\approx \beta$ by fitting $\Delta u$, $\Delta_3$ \cite{PDG}
as well as $\bar u-\bar d$, $\bar u/\bar d$ \cite{e866} leading to
$a=0.13$, $\zeta=-0.10$, $\alpha=\beta=0.45$ as the best fit
values. The parameters so obtained have been used to calculate the
spin polarization functions and the quatk distribution functions.
The calculated quantities pertaining to spin polarization
functions have been corrected by including the gluon polarization
effects \cite{cheng,ab} and symmetry breaking effects
\cite{cheng}. Similarly, the quark distribution functions have
been corrected by including the symmetry breaking effects. The
orbital angular momentum contributions to magnetic moment  are
characterized by the parameters of $\chi$CQM as well as the masses
of the GBs. For the $u$ and $d$ quarks, we have used their most
widely accepted values in hadron spectroscopy
\cite{cheng,yaoubook}, for example, $M_u=M_d=330$ MeV. For
evaluating the contribution of GBs, we have used its on mass shell
value in accordance with several other similar calculations
\cite{{mpi1}}.

In Tables \ref{spin}-\ref{quark}, we have presented the results of
our calculations pertaining to the strangeness dependent
parameters in $\chi$CQM$_{{\rm config}}$. For comparison sake, we
have also given the corresponding quantities in CQM. To begin
with, we first discuss the quality of fit pertaining to the spin
polarization functions. In Table \ref{spin}, we have presented the
strangeness incorporating spin polarization functions and the weak
axial vector couplings. Using $\Delta u$, $\Delta_3$ along with
$\bar u-\bar d$, $\bar u/\bar d$ from Table \ref{quark} as inputs,
we find that we are able to achieve an excellent fit in the case
of spin polarization functions and the weak axial vector
couplings. In particular, the agreement in terms of the magnitude
as well as the sign in the case of $\Delta s$ is in good agreement
with the latest data \cite{smc,adams}.  An excellent agreement in
the case of $\Delta_8$ and $\Delta_0$, which receives contribution
from $\Delta s$ also, not only justify the success of
$\chi$CQM$_{{\rm config}}$ but also strengthen our conclusion
regarding $\Delta s$. Similarly, the excellent agreement obtained
in the case of the ratio $F/D$ again reinforces our conclusion
that $\chi$CQM$_{{\rm config}}$ is able to generate qualitatively
as well as quantitatively the requisite amount of strangeness in
the nucleon.

In Table \ref{mag}, we have presented the strangeness spin and
orbital contributions pertaining to the magnetic moment of the
nucleon and  $\Delta$ baryons. From the Table one finds that the
present result for the strangeness contribution to the magnetic
moment of proton looks to be in agreement with the most recent
HAPPEX results \cite{happex} as well as with the lattice QCD
calculations \cite{mup-aw}. On closer examination of the results,
several interesting points emerge. The strangeness contribution to
the magnetic moment is coming from spin and orbital angular
momentum of the ``quark sea'' with opposite signs. These
contributions are fairly significant and they cancel in the right
direction to give the right sign and magnitude to $\mu(p)^s$, For
example, the spin contribution in this case is $-0.09 \mu_N$ and
the contribution coming from the orbital angular momentum is $0.05
\mu_N$. These contributions cancel to give $-0.03 \mu_N$ which is
very close to the observed HAPPEX results ($-0.038 \pm 0.042
\mu_N$). Interestingly, in the case of $\mu(n)^s$ the magnetic
moment is dominated by the orbital part. Therefore, an observation
of this would not only justify the Cheng-Li mechanism \cite{cheng}
but would also suggest that the chiral fluctuations is able to
generate the appropriate amount of strangeness in the nucleon. For
the sake of completeness, we have also presented the results of
$\mu(\Delta^{++})^s$, $\mu(\Delta^{+})^s$, $\mu(\Delta^{o})^s$,
$\mu(\Delta^{-})^s$ and here also we find that there is a
substantial contribution from spin and orbital angular momentum.
In general, one can find that whenever there is an excess of $d$
quarks the orbital part dominates, whereas when we have an excess
of $u$ quarks, the spin polarization dominates.

After finding that the  $\chi$CQM$_{{\rm config}}$ is able to give
an excellent account of the spin dependent polarization functions,
in Table \ref{quark}, we have presented the results of quark
distribution functions having implications for strangeness in the
nucleon. In line with the success of $\chi$CQM$_{{\rm config}}$ in
describing the spin dependent polarization functions, in this case
also we are able to give an excellent account of most of the
measured values. The agreement in the case of $\frac{2s}{u+d}$ and
$\frac{f_3}{f_8}$ indicates that, in the $\chi$CQM, we are able to
generate the right amount of strange quarks through chiral
fluctuation. A refinement in the case of the strangeness dependent
quark ratio $\frac{2s}{\bar u+\bar d}$ would have important
implications for the basic tenets of $\chi$CQM. The observed
result for the case of $f_s$ in the present case also indicates
that the strange sea quarks play a significant role in the
nucleon. This is in agreement with the observations of other
authors \cite{cheng,gasser,scadron}.

To summarize, the  $\chi$CQM$_{{\rm config}}$ is able to provide a
excellent description of the spin dependent polarization functions
and quark distribution functions having implications for
strangeness in the nucleon. It is able to give a quantitative
description of the important parameters such as $\Delta s$, the
weak axial vector couplings $\Delta_8$ and $\Delta_0$, strangeness
contribution to the magnetic moment $\mu(p)^s$, the strange quark
flavor fraction $f_s$, the strangeness dependent ratios $\frac{2
\bar s}{u+d}$ and $\frac{f_3}{f_8}$ etc.. In the case of
$\mu(p)^s$, our result is in full agreement with the latest
measurement as well as with the lattice QCD calculations. In
conclusion we would like to state that at the leading order
constituent quarks and the weakly interacting Goldstone bosons
constitute the appropriate degrees of freedom in the
nonperturbative regime of QCD and the ``quark sea'' generation in
the $\chi$CQM$_{{\rm config}}$ through the chiral fluctuation is
the key in understanding the strangeness content of the nucleon.

 \vskip .2cm
 {\bf ACKNOWLEDGMENTS}\\
H.D. would like to thank DST (Fast Track Scheme), Government of
India, for financial support and the chairman, Department of
Physics, for providing facilities to work in the department.

%\pagebreak

\begin{table}
\begin{center}
\begin{tabular}{cccc} \hline \hline
Parameter & Data  & CQM & $\chi$CQM$_{{\rm config}}$ \\   \hline

$\Delta u^*$ & 0.85 $\pm$ 0.05 \cite{smc} & 1.333   & 0.867
\\ $\Delta d$ & $-$0.41  $\pm$ 0.05 \cite{smc}   & $-$0.333 &
    $-$0.392 \\

$\Delta s$ & $-0.10 \pm 0.04 $ \cite{smc}   &   0 &
 $-0.08$
\\
& $-0.07 \pm 0.03 $ \cite{adams}   &        & \\

$\Delta_3^*$ & 1.267 $\pm$ 0.0035 \cite{PDG} & 1.666 &
 1.267\\ $\Delta_8$ & $0.58 \pm 0.025$ \cite{PDG} & 1
 & 0.59  \\ $\Delta_0$ & $0.19 \pm 0.025$ \cite{PDG} &
0.50     & 0.19
\\ $F/D$ & $0.575 \pm 0.016$ \cite{PDG} &
0.673     & 0.589
\\
\hline \hline {\small $*$ Input parameters} &&&

\end{tabular}
\end{center}
\caption{The calculated values of the strangeness dependent spin
polarization functions and weak axial vector couplings in the CQM
and $\chi$CQM$_{{\rm config}}$.}
 \label{spin}
\end{table}

\begin{table}
\begin{center}
\begin{tabular}{cccc} \hline \hline
Parameter & Data  & CQM & $\chi$CQM$_{{\rm config}}$ \\   \hline

$\mu(p)^s_{{\rm spin}}$, $\mu(p)^s_{{\rm orbit}}$ & $-$ &0, 0 &
$-0.09$, 0.05 \\ $\mu(p)^s$ & $ -0.36 \pm 0.20$\cite{sample}  & 0
& $-0.03$
\\ & $-0.038 \pm 0.042$
\cite{happex}   &     &  \\

$\mu(n)^s_{{\rm spin}}$, $\mu(n)^s_{{\rm orbit}}$ & $-$ &0, 0
 & 0.06, $-0.09$ \\ $\mu(n)^s$ & $-$& 0 &  $-0.03$
\\

$\mu(\Delta^{++})^s_{{\rm spin}}$, $\mu(\Delta^{++})^s_{{\rm
orbit}}$ & $-$ & 0, 0  & $-0.29$, 0.18  \\ $\mu(\Delta^{++})^s$ &
$-$ & 0 &
 $-0.11$ \\

$\mu(\Delta^{+})^s_{{\rm spin}}$, $\mu(\Delta^{+})^s_{{\rm
orbit}}$ & $-$ & 0, 0 & $-0.14$,  0.11 \\ $\mu(\Delta^{+})^s$ &
$-$ & 0 & $-0.03$ \\

$\mu(\Delta^{o})^s_{{\rm spin}}$, $\mu(\Delta^{o})^s_{{\rm
orbit}}$ & $-$ & 0, 0  &   $-0.04$, $-0.03$ \\ $\mu(\Delta^{o})^s$
& $-$ & 0 & $-0.07$ \\

$\mu(\Delta^{-})^s_{{\rm spin}}$, $\mu(\Delta^{-})^s_{{\rm
orbit}}$ & $-$ & 0, 0 & $-0.09$, 0.15 \\ $\mu(\Delta^{-})^s$ & $-$
& 0 & $0.06$
\\  \hline \hline

\end{tabular}
\end{center}
\caption{The calculated values of the strangeness contribution to
the magnetic moment of nucleon and $\Delta$ decuplet baryons in
the CQM and $\chi$CQM$_{{\rm config}}$.}
 \label{mag}
\end{table}

\begin{table}
\begin{center}
\begin{tabular}{cccc} \hline \hline
Parameter & Data  & CQM & $\chi$CQM$_{{\rm config}}$ \\   \hline

$\bar s$ & $-$   &   0 & 0.10 \\

$\bar u-\bar d^*$ & $-0.118 \pm$ 0.015 \cite{e866} & 0   &
 $-0.118$
\\

$\bar u/\bar d^*$ & 0.67 $\pm$ 0.06 \cite{e866}  & 0
 & 0.66\\

$\frac{2 \bar s}{u+d}$ & 0.099$^{+0.009}_{0.006}$ \cite{ao} & 0 &
 0.09
\\

$\frac{2 \bar s}{\bar u+\bar d}$ & 0.477$^{+0.063}_{0.053}$
\cite{ao} & 0  &  0.44
\\

$f_s$ &  0.10 $\pm$ 0.06 \cite{ao}  &  $-$ & $0.08$
\\
$f_3$  &$-$ & $-$  & 0.21\\

$f_8$  &$-$ & $-$ & 1.03\\ $f_3/f_8$ & 0.21 $\pm$ 0.05 \cite{ao} &
0.33 &   0.20 \\ \hline \hline {\small $*$ Input parameters} &&&

\end{tabular}
\end{center}
 \caption{The calculated values of the strangeness dependent quark flavor
distribution functions and related parameters in the CQM and
$\chi$CQM$_{{\rm config}}$.} \label{quark}
\end{table}

\end{document}